\documentclass[twocolumn,trackchanges]{aastex631}

\shorttitle{Milky Way galaxies-analogues}
\shortauthors{Vavilova et al.}
\graphicspath{{./}{figures/}}

\begin{document}

\title{AN ADVANCED APPROACH FOR DEFINITION OF\\ THE "MILKY WAY GALAXIES-ANALOGUES"}

\author[0000-0002-5343-1408]{Iryna B. Vavilova}
\affiliation{Main Astronomical Observatory, National Academy of Sciences of Ukraine, \\
27 Akademik Zabolotnyi Street, Kyiv, 03143, Ukraine}

\author{Petr M. Fedorov}
\affiliation{Institute of Astronomy, V,N, Karazin  National University of Kharkiv,\\
4 Svoboda Sq., Kharkiv, 61022, Ukraine}
  
\author{Daria V. Dobrycheva}
\affiliation{Main Astronomical Observatory, National Academy of Sciences of Ukraine, \\
27 Akademik Zabolotnyi Street, Kyiv, 03143, Ukraine}

\author{Olga Sergijenko}
\affiliation{Main Astronomical Observatory, National Academy of Sciences of Ukraine, \\
27 Akademik Zabolotnyi Street, Kyiv, 03143, Ukraine}

\author{Anatolii A. Vasylenko}
\affiliation{Main Astronomical Observatory, National Academy of Sciences of Ukraine, \\
27 Akademik Zabolotnyi Street, Kyiv, 03143, Ukraine}

\author{Artem M. Dmytrenko}
\affiliation{Institute of Astronomy, V,N, Karazin  National University of Kharkiv,\\
4 Svoboda Sq., Kharkiv, 61022, Ukraine}

\author{Vlad P. Khramtsov}
\affiliation{Institute of Astronomy, V,N, Karazin  National University of Kharkiv,\\
4 Svoboda Sq., Kharkiv, 61022, Ukraine}

\author{Olena V. Kompaniiets}
\affiliation{Main Astronomical Observatory, National Academy of Sciences of Ukraine, \\
27 Akademik Zabolotnyi Street, Kyiv, 03143, Ukraine}



\begin{abstract}

Our Galaxy — the Milky Way — has certain features of the structure and evolution. The morphological, photometric, kinematic, and
chemodynamical properties are usually considered in the search for the Milky Way galaxies-analogues (MWAs). The discovery of MWA galaxies with a larger number of simultaneous selection parameters and more stringent constraints on a given parameter yields
a sample of MWA galaxies with properties closer to the true properties of the Milky Way. So, in general, such MW parameters as the
morphological type, luminosity, color indices, structural parameters (size, bar, bulge, thin and thick disks, inner ring, halo), bulge-
to-total ratio, stellar mass, star formation rate, metallicity, and rotation velocity were used in various combinations for comparison
with other galaxies. However, the offset of some MW features in the multi-parameter space of MWAs features should be significant.

The paper aims to give a brief overview of the problematics and to present our approach for studying Milky Way and MWAs matching characteristics (this project is supported by the National Research Fund of Ukraine). We propose to enlarge as much as possible
the number of Milky Way features and compile various samples of MWAs in our co-moving cosmological volume for their further
optimization. Such features can include 3D-kinematics of star’s movement in certain regions, low oxygen content on the periphery, low nuclear activity, and the lack of significant merging over the past 10 Gyrs (isolation criterion). This approach will make it possible to widely formulate the necessary and sufficient conditions for the detection of MWA galaxies as well as to reveal other MW multiwavelength features.
\end{abstract}

\keywords{Galactic and extragalactic astronomy --- Galactic morphology --- Active galactic nuclei --- Milky Way --- Stellar kinematics --- Cosmological evolution}


\section{Introduction} \label{sec:intro}

The morphological, kinematic, and multiwavelength properties of the Milky Way (MW), the structure of spiral arms, the chemical evolution, the low activity of a supermassive black hole, the cosmological origin, and the placement of the Milky Way with neighboring galaxies in the cosmic web are essential questions of modern astrophysics. 

Is the Milky Way a typical giant spiral galaxy really, and if not, how is it different, and how many the MW-like galaxies are? The fact that the Milky Way has general ratios between various parameters, which are characterized for the spiral galaxies, suggests that the MW is not highly unusual among galaxies. So, the anthropic principle is true, at least as a first-order approximation. This justifies the selection of Milky Way galaxies-analogues (MWAs) using any MW parameters (stellar mass, luminosity, star formation rate, bulge-to-disk ratio, disk scale length, rotation velocity, and morphology are usually considered). However, the offset of the MW in some multi-parameter MWAs space can be significant, i.e. the MW has some individual features. 

In the earlier works, only the obvious morphological and photometric parameters were to be taken into account as the MW indicators for MWAs. For example, in the seminal article for this research field, \cite{deVaucouleurs1978} calculated isophotal $R_{25}$ = 11.5 kpc and effective $R_{eff}$ = 5.1 kpc radii as optical scale lengths, inner ring diameter $D(r)$ = 6 kpc, luminosity values ($M_{T}(B)$ = -20.1, class II, etc.), color $(B-V)_{T}$ = 0.53 and hydrogen indices in a frame of the two-component MW structure model (spheroidal for bulge and exponential for disk). It allowed them to identify four nearby galaxies NGC 1073, NGC 4303, NGC 5921, and NGC 6744 as possible MWAs. 

\cite{Mutch2011} compiled a list of MWAs using the stellar mass and the structural parameter as the selection criteria, where the latter parameter corresponds to the profile of the de Vaucouleurs brightness distribution. \cite{Licquia2015} selected an MWA sample from the SDSS-III DR8, forming it on the stellar mass and current star formation rate of these galaxies. \cite{Boardman2020a} presented a sample of 62 MWA galaxies from the MaNGA with selection criteria as the stellar masses and bulge-to-disk ratio. We note that the selection of MWA galaxies using only two MW characteristics is a typical approach and consistent with the anthropic principle that the Milky Way is not unusual among other galaxies. We should also take into account that our Galaxy has stable rotation periods: for general spiral pattern of 220–360 Myr \citep{Gerhard2011}, for bar pattern of 160–180 Myr \citep{Shen2020}, and for Sun's Galactic position of 212 Myr \citep{Reid2019}.

The discovery of MWA galaxies with a larger number of simultaneous selection parameters as well as more stringent constraints on a given parameter yields a sample of MWA galaxies with properties that are closer to the true features of the Milky Way. So, in general, such MW parameters as the morphological type, luminosity, color indices, structural parameters (size, bar, bulge, disk), bulge-to-total ratio, stellar mass, and rotation velocity were used in various combinations for comparison with other galaxies. 

On the other hand, when we use more selection criteria we add little or no MWA galaxies \citep{Boardman2020b}. For example, \citep{Fraser2019} found only 176 MWA galaxies among over a million galaxies of the DR7 SDSS selecting them by three parameters: stellar mass $M_{*}$, morphology, and bulge-to-disk ratio and using Galaxy Zoo morphological classification.  They concluded that the Milky Way is a galaxy with a low star formation rate log $({SFR}_{MW}/{M}_{Sun} yr^{-1})$=0.22 but it is not unusual when compared to similar galaxies. \cite{Boardman2020a} did not find MWA galaxies in the MaNGA survey when they combined four parameters: stellar mass, star-formation rate, bulge-to-disk ratio, and disk scale length. \cite{Tuntipong2024} used four selection parameters (stellar mass $M_{*}$, star formation rate ($SFR$), bulge-to-total ratio ($B/T$), and disk effective radius ($R_{eff}$) for identifying MWAs in the SAMI Galaxy Survey. Combinations of all the parameters allowed them to find 10 MWAs in the GAMA and Cluster regions of the SAMI survey and to outline that $B/T$ is the least important out of them. 

Pilyugin et al. in a series of works \citep{Pilyugin2014, Pilyugin2019, Pilyugin2023} developed the newest approach for the MWAs selection. The oxygen abundance characterizes an astration level (a fraction of matter converted into a star) and, consequently, is an indicator of how far a galaxy moved in its (chemical) evolution. They considered a sample of about 500 galaxies from the MaNGA survey with selection criteria for three parameters (stellar mass $M_{*}$, optical radius $R_{25}$, rotation velocity $V_{rot}$) to determine the oxygen abundance in the center $(O/H)_{0}$ and at the isophotal radius $(O/H)_{R25}$. They found that the $(O/H)_{R25}$ in the Milky Way is appreciably lower than in other galaxies with similar $(O/H)_{0}$. So, they revealed that the most prominent feature of our Galaxy is the low metallicity at the periphery and identified four galaxies (NGC 3521, NGC 4651, NGC 2903, and MaNGA galaxy M-8341-09101) that can be considered as Milky Way twins. 

Such results suggest that some MW parameter(s) can be unusual or their combination can be rare. In this sense, we propose an advanced approach to find MWAs pointing out as much as possible indicators of MW features (Section \ref{sec:concept}. It allows to enlarge selection criteria for MWA definition as well as look wider at how manner MW properties look for the outside extragalactic observer. The aim of our article is to present the project ''The Milky Way galaxies-analogues" (2024-2026) supported by the National Research Fund of Ukraine. 

\section{Concept of an advanced approach} \label{sec:concept}

 Being a typical barred spiral galaxy the Milky Way has several meaningful observational features of its evolution. In our opinion, using multiple selection criteria for finding MWAs is a more effective approach, which enlarges the common picture of MW and NWAs' physical properties, diminishes potential biases in MWA studies, and enriches our knowledge for identifying and optimizing the necessary/sufficient conditions for the definition of MWAs galaxies. 
 
 We briefly clarify and discuss below the MW parameters and observational features both mostly considered and those we would like to propose for MWAs search.

 \subsection{Morphological and photometric parameters, scale length of strictures, metallicity, stellar mass, star formation rate}
 
\textbf{Morphological type} of our Galaxy was determined as SAB(rs)bc by \cite{deVaucouleurs1978}. The SBc type (T=4) is usually considered for MWAs study.

\textbf{Main structural parameters and metallicity}: giant disk galaxy having four  (two ?) spiral arms with twist angles, bar, bulge, inner ring, and halo. The thickness of thin and thick disks are usually considered as 220–450 pc and 2.6 kpc, respectively \citep{Bland2016}.

\textbf{Metallicity} as a key parameter of the chemodynamical evolution reveals many factors, which define the MW structural parameters. 
\cite{Hammer2007} found a systematic offset in the position of the Milky Way in the parameter space within 1$\sigma$ for all Tully-Fisher ratios, which shows a significant deficiency in stellar mass, angular momentum, disk radius, and $Fe/H$ in the stars in periphery region at a given $V_{rot}$. In contrast, \cite{McGaugh2016} suggests that the Milky Way is a normal spiral galaxy obeying the Tully–Fisher and the size-mass relation. \cite{Licquia2016} examine the three-dimensional diagram ($V_{rot}$ – luminosity – radius) and found that the Milky Way lies farther from this relation than around 90 \% of other spiral galaxies, yielding evidence that MW is extremely compact for its rotation velocity and luminosity possessing a cold bar \citep{Quillen1996}. The recent studies by \cite{Queiroz2021} with the GAIA data highlight the differences of the stellar population in the MW bar and bulge.  

The chemical properties of MW are not typical in several aspects. On the one hand, the MW is one of the most oxygen-rich spiral galaxies in the sense that the metallicity in the center is close to the maximum attainable value \citep{Pilyugin2007}. On the other hand, the oxygen abundance along the optical radius is noticeably lower than in galaxies with a similar central metallicity \citep{Pilyugin2023}. At the same time, MW has a very steep metallicity gradient compared to most giant spiral galaxies, in which the change in oxygen excess along the optical radius is quite small \citep{Pilyugin2024}.

\cite{Chandra2023} considered the MW three-phase chemodynamical evolution exploring a sample of 10 million red giant stars with low-resolution Gaia XP spectra and operating with angular momentum as a function of metallicity. They compared it with those of MWAs from the Illustris(TNG50) cosmological simulation taking into account these three MW evolutionary phases: the disordered protogalaxy, the kinematically hot old disk, and the kinematically cold young disk. They proposed three physical mechanisms for explanation (spinup, merger, and cooldown), which satisfies conditions of the Gaia-Sausage-Enceladus (GSE) last major merger with our Galaxy at $z \approx$ 2. In the frame of this three-phase scenario, \cite{Semenov2024} proposed an explanation of yet one MW feature: the MW disk was formed quite early, within the first few billion years of its evolution. It is consistent with the overall population of MWA-mass disk galaxies. 

Among \textbf{other MW chemodynamical features}, which were compared with MWAs obtained in TNG50 cosmological simulations we note recent fundamental research by \cite{Rix2024}. Using Gaia XP spectra they found a universal feature for MW and MWAs: their extremely metal-rich giant stars ($M/H_{XP}$ > 0) are mostly concentrated in a \textbf{compact central dynamically hot knot} with $R <$ 1.5 kpc. Taking into account that MW metal-poor stars are also concentrated in the central few kiloparsecs region, future studies with the SDSS-V as write these authors will allowing to estimate the stellar population more precisely. Another ''side" of our Galaxy, the \textbf{halo} at 10-80 kpc, was studied by \cite{Han2024} with the H3 Survey data. They found the strong kinematic asymmetries of distributions and, consequently, cold and hot kinematically fractions of stars with radial velocity dispersions 70 km/s and 160 km/s, respectively.

As for the \textbf{inner ring} feature, \citep{Wylie2022} in their recent meticulous work with a sample of APOGEE DR16 inner Galaxy stars studied the outer bar region. They considered the orbits of stars in the ''state of the art bar-bulge potential with a slow pattern speed" constructing the maps of their metallicity [Fe/H], density, and ages. They conclude that MW inner ring-like structure is on average middle-age and has a metal-rich gradient along the bar's major axis. Its location is between the planar bar and corotation radius. 

The following \textbf{photometric parameters} are usually taken into account for MWA search: luminosity class (II); isophotal diameter $D_{25}$ = 26.8 kpc \citep{Goodwin1998}, where the isophotal radius is usually adopted as $R_{25}$ = 12 kpc; the stellar disk up to 1.35 kpc \citep{Rix2013}. To learn the ''Maps of the Milky Way based on Gaia DR3 from 10 pc to 200 kpc" one can go to the next url: \url{https://gruze.org/posters_dr3/}.

The \textbf{total mass} of our Galaxy has various estimates depending on methods and entire MW region for this estimate: from 8.5$\times$10$^{11}$ $M_{Sun}$ \citep{Penarrubia2014}
up to 1.4$\times$10$^{12}$ $M_{Sun}$ \citep{Grand2019}. The \textbf{virial mass} at Galactocentric distance less than 21.1 kpc is $M_{vir}$ = 0.2 $\times$ 10$^{12} M_{Sun}$ (see, for example, \cite{Watkins2019} for estimation by halo globular clusters motion). The \textbf{stellar mass} is $M_{*}$$= 5\times$ $10^{10}M_{Sun}$ (log$M_{*}$ = 10.7) and  \textbf{star formation rate} $SFR$ = (1.78 $\pm$ 0.36) $M_{Sun}$ yr$^{-1}$ with taking the linear scale length ($B/T$, $R_{eff}$) into consideration \citep{Tuntipong2024}; number of stars $N_{*}$ = (1–4) $\times 10^{11}$ when the disk stars were detected with Gaia DR2 even beyond 25 kpc from the MW center. The \textbf{dark matter density} at Sun's position $M_{DM}$ = 0.0088 $M_{Sun}$ pc$^{-3}$ \citep{Kafle2014} but a dark matter area may extend up to $\approx$ 600 kpc \citep{Deason2020}. 

\subsection{Nuclear activity, supermassive black hole, 3D-kinematics of stars}

We propose to search for such a parameter of the Milky Way in the parameter space, which shows the maximum deviation from the corresponding relation for spiral galaxies. This parameter is used as the main criterion for the selection of MWA galaxies. Because the search for MWA galaxies is feasible using any parameters of the MW, we propose to increase their number both for the search for MWAs and for the specification of the MWAs properties. 

In most of the research, as you see, the MW galaxies-analogues were selected based on two/three parameters: stellar mass and some additional parameter, usually bulge-to-disk (bulge-to-total) ratio, the star formation rate. In addition to these MW features, we would like to highlight a weak nuclear activity and low mass of the supermassive black hole (SMBH), 3D-kinematics for the rotation velocity, isolation criteria, and several known multiwavelrngth properties. 

\textit{3D-kinematics of star's movement} can serve as an indicator for MWAs search. In series of works, \cite{Fedorov2021, Fedorov2023}, \cite{Dmytrenko2023}, \cite{Denyshchenko2024}  investigated region of the Milky Way in coordinate ranges 120$^\circ$ $<$ 
 $\theta$ $<$ 240$^\circ$, 0 kpc $<$ $R$ $<$ 16 kpc, -1 kpc $<$ $Z$ $<$ 1 kpc with Gaia EDR3 using samples of red giants and sub-giants whose centroids are in the MW plane. For the first time, these authors derived the dependence of  the centroid's kinematic parameters on the Galactocentric coordinates as well as the parameters for rotational velocity $\partial$$V_{R}$/$\partial$$\theta$ and $\partial$V$\theta$/$\partial$$\theta$. 

Our approach includes the task of investigating the 3D-kinematics of a large part of the MW based on GAIA DR3 data within the Ogorodnikov-Milne model and using the determined strain and rotation rate tensors to establish the spiral pattern of the MW. The studied galactic space will be bounded by the Galactocentric coordinates 4 kpc $<$ $R$ $<$ 14 kpc, 140$^\circ$ $<$ $\theta$ $<$ 220$^\circ$ and -3 kpc $<$ $Z$ $<$ 3 kpc. This is the region dominated by older stars, which are more evenly distributed than younger blue stars. Using the obtained components of the spatial velocities of the centroids and kinematic parameters we will able to construct $V_{rot}(R)$ and their slope within this region and to determine the parameters of the spiral arms including the coordinates of the vertices of various star regions. These results can serve as indicators for applying machine learning in tasks for search of MWAs with the kinematically cold rotating disk.

We remind that our Galaxy possesses both a weak \textbf{nuclear activity} and small mass of the \textbf{supermassive black hole}: $M_{SMBH}$ = 4.61 $\times$ 10$^{6} M_{Sun}$ \citep{Ghez2008, Becerra-Vergara2021}. The central object Sgr A* usually shows quiescent state, but sometimes does show rapid outbursts or flares of radiation (see, e.g., \cite{Genzel2010, Dodds2011}, and references therein). This is the case of a low-luminosity galactic nucleus, radiating at $\approx$10$^{–8}$ of the Eddington level. 
In the such regime of activity, the MW core has no typical AGN-like gas-dusty torus but has so-called ''circumnuclear disk” (CND) as a torus-like dusty-molecular gas around Sgr A$\star$ extending from $\approx$ 1 pc to $\approx$ 5 pc \citep{Etxaluze2011, Lau2013, Tsuboi2018}.

\subsection{Isolation criterion}
We adopt as a working hypothesis that the features of the MW are caused by its evolution without major merging for the last 10 Gyr. We consider the isolation criterion of MWA galaxies on the scales of neighboring galaxies \citep{Tully1987} to study the role of satellites in the evolution of MWA galaxies. The Milky Way can be considered as an isolated galaxy during the long time of its evolution. The results of the high-resolution N-body simulations of last major merger \citep{Naidu2021} allowed in particular to determine both the orbital parameters of merging with two density profile breaks at $\approx$ 15-18 kpc and 30 kpc as well as distribution of stellar and dark matter mass between GSE and Milky Way. 

What is about minor mergers? What is the influence of the Magellanic Clouds (the gas reservoirs) on the MW evolution? \citep{vandenBergh2006} assumes that the Magellanic Clouds may be interlopers from a remote part of the Local Group rather than true satellites of the Milky Way, i.e. the Large Magellanic Cloud (LMC) is on its first approach to the MW. \cite{Font2021} and \cite{Jones2024} investigated the significance of satellite effects and analyzed star formation rates in galaxy systems similar to the MW-like satellite system using the projected satellite–host galaxy distances. Exploiting the data from SAGA II catalog, CFHT and VLA observations, \cite{Jones2024} defined eight such MW-like systems when a satellite is undergoing ram pressure stripping in process of collision with circumgalactic medium of its host galaxy.   

What is for the future collision of the Milky Way with Andromeda galaxy (\citep{Dubinski1996,  vandenMarel2019}) in around 5 Gyr? \citep{Sawala2024} used the Gaia and HST observational data to determine the dynamical process of merging the MW-M31 system. These authors predicted that M33 and LMC as other members of the Local Group are able to make this merger less likely because the LMC orbit runs perpendicular to the MW-M31 system orbit. Moreover, they found that existing uncertainties in the present kinematic and dynamic (masses) data for Local Group galaxies give 50 \%-50 \% probability of MW-M31 merger during the next 10 Gyr. The certain role can plays not only the correct distance moduli determination (see, for example, \cite{Elyiv2020}), the MW-M31 orbital geometry (\citep{Banik2022}), but the position of Local Void lying adjacent to the Local Group (\citep{Tully1987, Lindner1995, Mazurenko2024}) and the MW moving away this void.

To study the role of interaction with neighboring (dwarf/normal) galaxies in the evolution of MWA galaxies, the isolation parameters are available for nearby galaxies in catalogs of isolated galaxies \citep{Sorgho2024}. For example, the isolated galaxies selected from 2MIG catalog exhibit \textit{multiwavelength properties}, which are characterized by a weak \textit{nuclear activity} as compare with galaxies in the dense environment (\citep{Pulatova2015, Melnyk2015, Volvach2011, Dobrycheva2018}) and faint luminocity in spectral ranges especially in radio- and X-ray ranges (\citep{Pulatova2023, Vasylenko2020}). 

For the most MW-like galaxy \textit{NGC 3521} (MW-twin in terms of baryon mass, rotation velocity, scaled disk length, metallicity), the observational data from the Ukrainian UTR-2 radio telescope (\citep{Konovalenko2016}) will be quite useful in decameter range both to have a full spectral energy distribution of the NGC 3521 and to explain the Galactic background radio emission, the North Polar Spur \citep{Miroshnichenko2009}. The archive of UTR-2 radio telescope contains large volumes of 24-hour survey data for 4-5 positions on inclination.

An interesting expected result to find the view of MW from the outside observer can be achieved using machine learning for classification by a range of photometric parameters (morphology details, optical radius, luminosity concentration index to the center, color indices, etc.) and image feature as the bar and bulge, structure of spiral arms, inclination angle, etc. (\citep{Vavilova2021, Khramtsov2022, Vavilova2022, Dobrycheva2023}). The obtained 3-D kinematics of the MW red-giant and sub-giant stars, and the parameters of multiwavelength radiation of MWAs could be considered as additional indicators.

The \textit{samples of candidates to the MW galaxies-analogues} should contain the MW features as much as possible. It allows optimizing the necessary and sufficient conditions for revealing MWAs. Thanks to this, the number of candidates for MWA galaxies will increase quantitatively, and their study will, in turn, allow us to understand the appearance and features of the MW as the extragalactic object. Cosmological simulations TNG50, in own turn, allow clarifying whether single/different evolutionary tracks lead to the MWA formation.

\section{Conclusion}

Our project for search MWAs includes several research areas: 3D kinematics of stars of the MW; selected multiwavelength properties of MWA galaxies, e.g. NGC 3521 as the most MW-like galaxy; the activity of the nuclei and SMBH masses of MWA galaxies; the gravitational coupling of selected MWA galaxies and the analysis of the significance of the influence of dwarf neighboring galaxies; cosmological simulations of the evolutionary tracks of MWA galaxies; search for the appearance of the MW for the outside observe by machine learning, which is based on the data of MWA-galaxies; the revealing for other features of the MW in comparison with the MWA galaxies.

The Milky Way galaxies-analogues provide an alternative insight into the various pathways that lead to the formation of disk galaxies with similar properties to the Milky Way. Such an approach will make it possible to widely formulate the necessary and sufficient conditions for the detection of MWA galaxies as well as to reveal MW multiwavelemgth features \citep{Vavilova2020}. 

In our cosmological comoving volume the MWAs can be identified in the process of solving the multi-parameter task of optimization of these quantitative and qualitative characteristics, which should be as much as possible similar to the MW features. This circumstance reflects the fact that available samples of MWAs contain the galaxies in the redshift range of the Local Universe. 

In this content, the Illustris(TNG50) simulations containing approximately 100 MWA galaxies by mass, could be used to define MWAs evolutionary tracks in order to discover whether there is a typical scenario of evolution that leads to the formation of MWA galaxies, or to estimate the probabilities of different scenarios of their formation. The obtained data will be exploited to select the most likely scenario (scenarios) for search of MWA galaxies at the higher redshifts.

Determining what our Galaxy (our big house) looks like from the outside is of great importance both directly to astrophysics and to the popularization of astronomy. Mankind has always been interested in whether our place of residence in the Universe is special, or whether there are other similar places. First, there was a search for planets near other stars, after the discovery of the first exoplanets, the search for terrestrial planets and planetary systems similar to the Solar System has began. The search for MWA galaxies is the next step on this path. 

\begin{acknowledgments}
The work is supported by the National Research Fund of Ukraine (Project No. 2023.03/0188).  
\end{acknowledgments}

\bibliography{MWA-ver1}{}
\bibliographystyle{aasjournal}

\end{document}